\documentclass[
journal=nalefd,
manuscript=letter,
layout=twocolumn
]{achemso}
\usepackage[utf8]{inputenc}

\usepackage{graphicx} 
\usepackage{amsmath}
\usepackage{hyperref} 
\usepackage{multirow}
\usepackage{array}

\usepackage{newtxtext}
\usepackage{xspace}

\usepackage{booktabs}
\usepackage[table]{xcolor} 
\usepackage{hyperref}
\hypersetup{colorlinks, 
	linkcolor={blue!75!black!80!yellow},
	citecolor={blue!75!black!80!yellow}, 
	urlcolor={blue!75!black!80!yellow}
	}

\newcommand{\hlcell}{\cellcolor{yellow!50}}

\renewcommand{\vec}[1]{\mathbf{#1}}
\newcommand{\sub}[1]{\ensuremath{_{\textrm{#1}}}}   \newcommand{\rlUnit}{$\times 10^{-16}~\Omega$m$^2$\xspace}

\newcommand{\CrAlC}{CrAl$_2$C\xspace}
\newcommand{\PtCoO}{PtCoO$_2$\xspace}
\newcommand{\PdCoO}{PdCoO$_2$\xspace}
\newcommand{\YCoB}{YCo$_3$B$_2$\xspace}
\newcommand{\ScCoB}{ScCo$_3$B$_2$\xspace}
\newcommand{\MoNi}{MoNi$_2$\xspace}
\newcommand{\CrNi}{CrNi$_2$\xspace}
\newcommand{\VNi}{VNi$_2$\xspace}
\newcommand{\VPt}{VPt$_2$\xspace}
\newcommand{\NiIr}{NiIr$_3$\xspace}

\newcommand{\RPIMSE}{Materials Science and Engineering, Rensselaer
Polytechnic Institute, Troy, NY 12180, USA}
\newcommand{\RPIPHYS}{Physics, Applied Physics and Astronomy, Rensselaer
Polytechnic Institute, Troy, NY 12180, USA}

\title{Ultralow Electron-Surface Scattering in Nanoscale Metals Leveraging Fermi Surface Anisotropy}

\author{Sushant Kumar}
\affiliation{\RPIMSE}

\author{Christian Multunas}
\affiliation{\RPIPHYS}

\author{Benjamin Defay}
\affiliation{\RPIPHYS}

\author{Daniel Gall}
\affiliation{\RPIMSE}

\author{Ravishankar Sundararaman}
\affiliation{\RPIMSE}
\affiliation{\RPIPHYS}
\email{sundar@rpi.edu}

\begin{document}

\noindent\textbf{TOC Image:}\\
\includegraphics[width=3.25in]{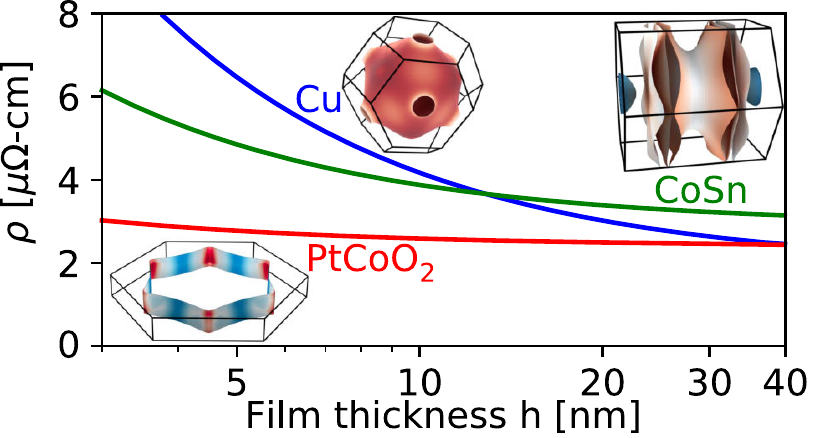}

\noindent\textbf{Abstract:}
Increasing resistivity of metal wires with reducing nanoscale dimensions is a major performance bottleneck of semiconductor computing technologies.
We show that metals with suitably anisotropic Fermi velocity distributions can strongly suppress electron scattering by surfaces and outperform isotropic conductors such as copper in nanoscale wires.
We derive a corresponding descriptor for the resistivity scaling of anisotropic conductors, screen thousands of metals using first-principles calculations of this descriptor and identify the most promising materials for nanoscale interconnects.
Previously-proposed layered conductors such as MAX phases and delafossites show promise in thin films, but not in narrow wires due to increased scattering from side walls.
We find that certain intermetallics (notably CoSn) and borides (such as \YCoB) with one-dimensionally anisotropic Fermi velocities are most promising for narrow wires.
Combined with first-principles electron-phonon scattering predictions, we show that the proposed materials exhibit 2-3$\times$ lower resistivity than copper at 5~nm wire dimensions.


\vspace{0.1in}\hrule\vspace{0.1in}
\setlength{\abovedisplayskip}{4pt}
\setlength{\belowdisplayskip}{4pt}

Miniaturization of semiconductor integrated circuits leads to higher transistor density and performance,\cite{salahuddin2018era, ferry2008nanowires, banerjee2001global, charles2005miniaturized} but computing performance beyond the 10~nm technology node is increasingly limited by $RC$ delays in the nanoscale copper wires interconnecting the transistors.\cite{salahuddin2018era,gall2020search, gall2021bulletin} The challenge stems from the dramatic increase of resistivity of metals when wire dimensions reduce below the electron mean free path $\lambda$ ($\approx 40$~nm for copper) due to surface and grain boundary scattering.\cite{gall2020search} 
Several strategies are being actively ivestigated to address this bottleneck, including using metallic nanowires \cite{xu2015copper, simbeck2012aluminum, lanzillo2014pressure, lanzillo2017ab}, doped multilayer-graphene-nanoribbons \cite{jiang2017intercalation}, two-dimensional metals\cite{hu2022van} and topological semimetals \cite{zhang2019ultrahigh,chen2020topological, han20211d} and insulators\cite{philip2016performance}. 
However, reliable interconnect materials that systematically outperform elemental metals like Cu remains a critical challenge.

Within the approximate semi-classical Fuchs-Sondheimer (F-S) \cite{fuchs1938math, sondheimer1952adv} and Mayadas-Shatzkes (M-S) models\cite{mayadas1970electrical} the resistivity of a polycrystalline square metal wire is
\begin{align}
\rho = \rho_0 
+ \rho_0 \lambda  \frac{3(1-p)}{4d}
+ \rho_0 \lambda \frac{3R}{2D(1-R)}
\label{eqn:FS}
\end{align}
where $\rho_0$ is the bulk resistivity of the metal, $d$ is the wire thickness and $D$ is the average grain diameter.
(For thin films, replace $4d$ in the denominator of the second term by $8h$, where $h$ is the film thickness.\cite{fuchs1938math, sondheimer1952adv})
Here, surface specularity $p$ and grain boundary reflectivity $R$ are typically used as phenomenological parameters to fit measured resistivities that exhibit the characteristic $1/d$ increase with reducing dimensions.\cite{gall2016electron, chen2018nial, chen2021intermetallics}

\begin{figure*}[t]
\centering
\includegraphics[width=\textwidth]{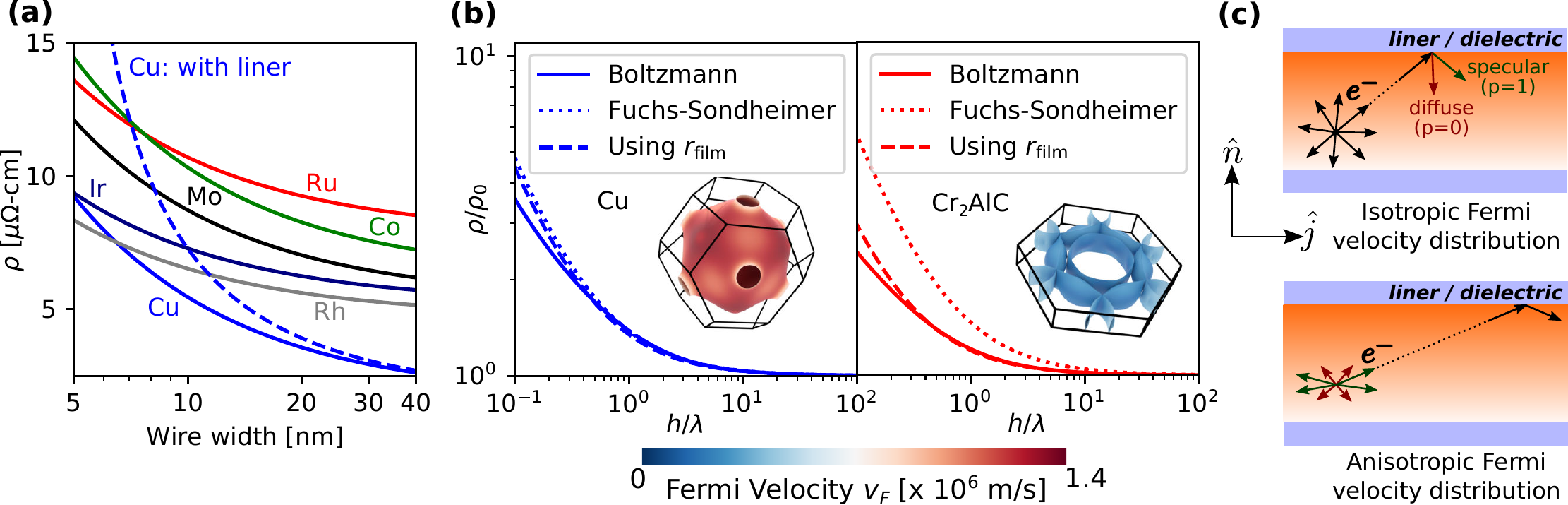}
\caption{(a) Resistivity increases with decreasing width of single-crystalline square wires due to increased surface scattering, shown here using the Fuchs-Sondheimer (F-S) model (Eq.~\ref{eqn:FS}) with $p=0$ and $\rho_0 \lambda$ calculated from first-principles.\cite{gall2016electron}
The need for a liner for Cu (assumed 2~nm thick) reduces conducting cross section and increases resistivity further.
(b) For isotropic conductors such as Cu, the F-S model is accurate compared to Boltzmann transport predictions for the increase of resistivity relative to bulk value $\rho_0$ with decreasing film thickness $h$ compared to mean-free-path $\lambda$, but it overestimates this effect for anisotropic conductors such as the \CrAlC MAX phase.
Replacing $\rho_0 \lambda$ with $r\sub{film}$ derived here (Eq.~\ref{eqn:r_film}) fixes this discrepancy.
(Insets show the corresponding Fermi surfaces.)
(c) In anisotropic conductors, electrons with small velocity components encounter the surface less frequently, leading to the slower resistiviy increase with reducing dimensions that we exploit here for new nanoscale interconnect materials.}
\label{fig:intro}
\end{figure*}

Figure~\ref{fig:intro}(a) shows the variation of resistivity as a function of square wire width for single crystals of several elemental metals as predicted by Eq.~\ref{eqn:FS}, assuming the worst-case diffuse limit $p = 0$ of surface scattering, using the bulk experimental resistivity $\rho_0$ and factor $\rho_0\lambda$ calculated from first-principles.\cite{gall2016electron}
(The final grain boundary term of Eq.~\ref{eqn:FS} vanishes with $D\to\infty$ for single crystals.)
Note that Rh, Ir and Mo have a higher $\rho_0$ than Cu, but their resistivity becomes lower than Cu for small enough wires because they have a smaller $\rho_0 \lambda$, which is a common prefactor in the resistivity increase due to both surface and grain boundary scattering.
Additionally, Cu must be surrounded by a liner material to promote adhesion and prevent diffusion of Cu atoms into the surrounding dielectric.
This reduces the cross section of Cu within the total space available for the wire,\cite{kaloyeros2000ultrathin, gall2020search} increasing the effective resistivity even more rapidly as shown in Figure~\ref{fig:intro}(a).

Consequently, interconnects in next-generation semiconductor devices require a material with low resistivity at nanoscale dimensions, in addition to being resistant to electromigration allowing thinner or no liners.
High-throughput screening using first-principles calculations can be invaluable in identifying promising materials, but resistivity at nanoscale dimensions is too computationally expensive to predict directly for thousands of materials.
Instead, $\rho_0 \lambda$ serves as a `resistivity scaling coefficient' within the F-S model that can be calculated rapidly from the Fermi velocities over the Fermi surface of the metal calculated using density-functional theory (DFT).\cite{gall2016electron}
Consequently, $\rho_0 \lambda$ has been used extensively as a descriptor to screen elemental metals,\cite{gall2016electron} intermetallics\cite{chen2018nial} and MAX phases (metallic carbides and nitrides).\cite{sankaran2021ab, zhang2021resistivity}
The measured resistivity increase in epitaxial films of many elemental metals and intermetallics agree reasonably with calculated  $\rho_0 \lambda$ values,\cite{gall2020search, chen2018nial, chen2021intermetallics} but they do not agree for highly anisotropic conductors.
For example, in the Ti$_4$SiC$_3$ MAX phase material, the calculated $\rho_0 \lambda$ is 5$\times$ larger than the value extracted from measured resistance of epitaxial films.\cite{zhang2021resistivity}
This overestimation of the resistivity increase of anisotropic materials indicates that high-throughput materials screening using $\rho_0\lambda$ as a descriptor of nanoscale resistivity may miss promising candidates.

In this Letter, we derive new resistivity scaling coefficients $r\sub{film}$ and $r\sub{wire}$ that capture the resistivity increase of films and wires of any material, fully accounting for anisotropy in excellent agreement with Boltzmann transport simulations.
With a comparable computational cost to $\rho_0\lambda$, we are able to calculate these coefficients from first-principles for thousands of metallic materials to find the most conductive metals at nanoscale dimensions.
We find promising candidates among several material classes including intermetallics, oxides and borides, which we investigate further using first-principles electron-phonon scattering simulations to identify the materials that can significantly outperform Cu in nanoscale wires.

\textbf{Anisotropic Conductance Descriptor:}
We begin by comparing the resistivity scaling for thin films of the nearly isotropic metal, Cu,\cite{gall2016electron, brown2016nonradiative} and the highly anisotropic layered MAX phase conductor, \CrAlC,\cite{ito2017electronic, sankaran2021ab, ouisse2017magnetotransport} as limiting cases of interest in Figure~\ref{fig:intro}(b) with the Fermi surfaces shown as insets.
The approximate F-S model predictions of $\rho/\rho_0$ agree with the Boltzmann transport simulations for Cu, except for extremely thin films with thickness $h \ll \lambda$, the electron mean-free path.
However, the F-S model strongly overestimates the resistivity increase in \CrAlC, for all $h$ including $h \gg \lambda$.

To understand this discrepancy, consider the conductivity of thin films from the Boltzmann transport equation in the relaxation time approximation,\cite{fuchs1938math, lucas1965electrical, zheng2017anisotropic}
\begin{multline}
\sigma(h) = \sum_b \int\sub{BZ} \frac{e^2 g_s d\vec{k}}{(2\pi)^3}
(-f'_0(\varepsilon_{\vec{k}b})) 
\left(\vec{v}_{\vec{k}b}\cdot\hat{j}\right)^2 \tau
\\
\times \left[
1 - \frac{|\vec{v}_{\vec{k}b}\cdot\hat{n}|\tau}{h}
\left(1 - \exp\frac{-h}{|\vec{v}_{\vec{k}b}\cdot\hat{n}|\tau}\right)
\right]
\label{eqn:BTE}
\end{multline}
where $\varepsilon_{\vec{k}b}$ and $\vec{v}_{\vec{k}b}$ are the electronic energies and velocities of band $b$ and wavevector $\vec{k}$ in the Brillouin zone BZ, $g_s=2$ is the spin degeneracy factor (neglecting spin-orbit coupling) and $\tau$ is the relaxation time.
Above, the derivative $-f'_0(\varepsilon_{\vec{k}b})$ of the Fermi-Dirac occupations restricts contributions to within a few $k_BT$ of the Fermi energy.

The terms on the first line of Eq.~\ref{eqn:BTE} depend only on velocities along the current direction $\hat{j}$ and capture the bulk contribution to conductivity.
The terms on the second line depend on velocities along the surface normal direction $\hat{n}$ and account for surface scattering (assuming the diffuse $p=0$ case here for simplicity).
Intuitively, this factor accounts for the fact that electrons with velocities nearly parallel to the surface encounter the surface much less frequently, and hence are less likely to be scattered, than those with a significant normal velocity component (Figure~\ref{fig:intro}(c)).
In contrast,
\begin{equation}
\frac{1}{\rho_0 \lambda} = \sum_b \int\sub{BZ}
\frac{e^2 g_s d\vec{k}}{(2\pi)^3} (-f'_0(\varepsilon_{\vec{k}b}))
\frac{\left(\vec{v}_{\vec{k}b}\cdot\hat{j}\right)^2}
{|\vec{v}_{\vec{k}b}|}
\label{eqn:rho_lambda}
\end{equation}
does not explicitly depend on velocities along $\hat{n}$ and misses this critical physical effect.
Consequently, $\rho_0 \lambda$ misses the advantage of small 
$|\vec{v}_{\vec{k}b}\cdot\hat{n}|$ in anisotropic conductors and overestimates the resistivity increase with reducing film thickness.

We therefore retain explicit dependence on velocities along the surface normal $\hat{n}$ and asymptotically expand Eq.~\ref{eqn:BTE}.
For large $h \gg |\vec{v}_{\vec{k}b}\cdot\hat{n}| \tau$, we can neglect the exponential in the final term of Eq.~\ref{eqn:BTE} to find $\sigma(h) \approx g_1(\hat{j}) \tau + g_2(\hat{j}, \hat{n})\tau^2/h$, with
\begin{multline}
g_1(\hat{j}) \equiv \sum_b \int\sub{BZ}
\frac{e^2 g_s d\vec{k}}{(2\pi)^3} (-f'_0(\varepsilon_{\vec{k}b}))
\left(\vec{v}_{\vec{k}b}\cdot\hat{j}\right)^2
\label{eqn:g1}
\end{multline}
and
\begin{multline}
g_2(\hat{j}, \hat{n}) \equiv \sum_b \int\sub{BZ}
\frac{e^2 g_s d\vec{k}}{(2\pi)^3} (-f'_0(\varepsilon_{\vec{k}b}))
\\ \times 
\left(\vec{v}_{\vec{k}b}\cdot\hat{j}\right)^2
|\vec{v}_{\vec{k}b}\cdot\hat{n}|.
\label{eqn:g2}
\end{multline}
From the above, the resistivity varies as $\rho(h) \approx \rho_0 + g_2(\hat{j},\hat{n})/(g_1(\hat{j})^2 h)$, which is equivalent to the F-S model for single-crystal thin films (Eq.~\ref{eqn:FS} with $4d\to 8h$ and $D\to\infty$), but with $\rho_0\lambda$ replaced by
\begin{equation}
r\sub{film} \equiv \frac{8g_2(\hat{j}, \hat{n})}{3g_1(\hat{j})^2}.
\label{eqn:r_film}
\end{equation}
Figure~\ref{fig:intro}(b) shows that using this new resistivity scaling coefficient, $r\sub{film}$, instead of $\rho_0\lambda$ in the F-S model agrees very well with the Boltzmann transport simulations for both isotropic Cu and anisotropic \CrAlC.
In fact, the only deviations are for $h \ll \lambda$, where the semi-classical Boltzmann transport equation is anyway no longer valid.
Therefore, $r\sub{film}$ is adequate as a descriptor of nanoscale resistivity increase for high-throughput screening of interconnect materials, regardless of the anisotropy of conduction.

Importantly, calculating $r\sub{film}$ is of comparable computational cost to $\rho_0\lambda$, just requiring two integrals of Fermi velocities over the Fermi surface in Eqs.~\ref{eqn:g1} and \ref{eqn:g2}, instead of single one.
We reiterate that the explicit dependence of $g_2(\hat{j}, \hat{n})$ on velocities along the surface normal is critical for capturing the effect of anisotropy.
Notice that $g_2(\hat{j}, \hat{n})$ and $r\sub{film}$ are not tensors and can depend sensitively on directions, even for a materials where symmetry requires tensors to be isotropic.
Even for cubic Cu, $\rho_0\lambda = 6.7$\rlUnit in any direction, while $r\sub{film}$ ranges from (6.1 to 7.2)\rlUnit depending on the directions of $\hat{j}$ and $\hat{n}$; only a perfectly spherical Fermi surface would lead to $r\sub{film} = \rho_0\lambda$ in all directions.
Among the elemental metals, cubic tungsten is an extreme case with $\rho_0\lambda = 8.1$\rlUnit in any direction, while $r\sub{film}$ ranges from (4.6 to 12.1)\rlUnit due to the highly directional velocities stemming from the shape of the Fermi surface.\cite{zheng2017anisotropic}
Due to this direction dependence, for each material, we find the combination of perpendicular directions, $\hat{j}$ and $\hat{n}$, that minimize $r\sub{film}$.

We can straightforwardly generalize the above results for thin films to the case of rectangular wires, where electrons can scatter from the side walls in addition to the top and bottom surfaces.
We find that replacing $\rho_0\lambda$ in the F-S model (Eq.~\ref{eqn:FS}) by
\begin{equation}
r\sub{wire} \equiv \frac{8\left(
g_2(\hat{j}, \hat{n}_1) w + g_2(\hat{j}, \hat{n}_2) h
\right)}{3 g_1(\hat{j})^2 (w+h)},
\label{eqn:r_wire}
\end{equation}
where $\hat{n}_1$ and $\hat{n}_2$ are the surface normals along the height $h$ and width $w$ directions respectively, analogously matches the asymptotic expansion of the Boltzmann transport solution for rectangular wires (see SI for details).
The additional constraint of side wall scattering leads to $r\sub{wire} \ge r\sub{film}$, with equality for a spherical Fermi surface.
For copper, $r\sub{film} = 6.1$\rlUnit only increases slightly to $r\sub{wire} = 6.2$\rlUnit for square wires ($w = h$).
However, for \CrAlC, $r\sub{film} = 3.4$\rlUnit increases substantially to $r\sub{wire} = 6.5$\rlUnit for square wires, as most of the advantage for films compared to $\rho_0\lambda = 8.0$\rlUnit seen in Figure~\ref{fig:intro}(b) is lost due to side-wall scattering for wires.
Consequently, to minimize $r\sub{wire}$ for optimal scaling for narrow wires, we need to find materials with velocities that are directional along a single $\hat{j}$ direction, with small components along both remaining perpendicular directions, $\hat{n}_1$ and $\hat{n}_2$.
Below, unless mentioned otherwise, we report $r\sub{wire}$ for square wires, and find the combination of mutually perpendicular $\hat{j}$, $\hat{n}_1$ and $\hat{n}_2$ that minimize $r\sub{wire}$.

\begin{figure*}[t]
\centering
\includegraphics[width=\textwidth]{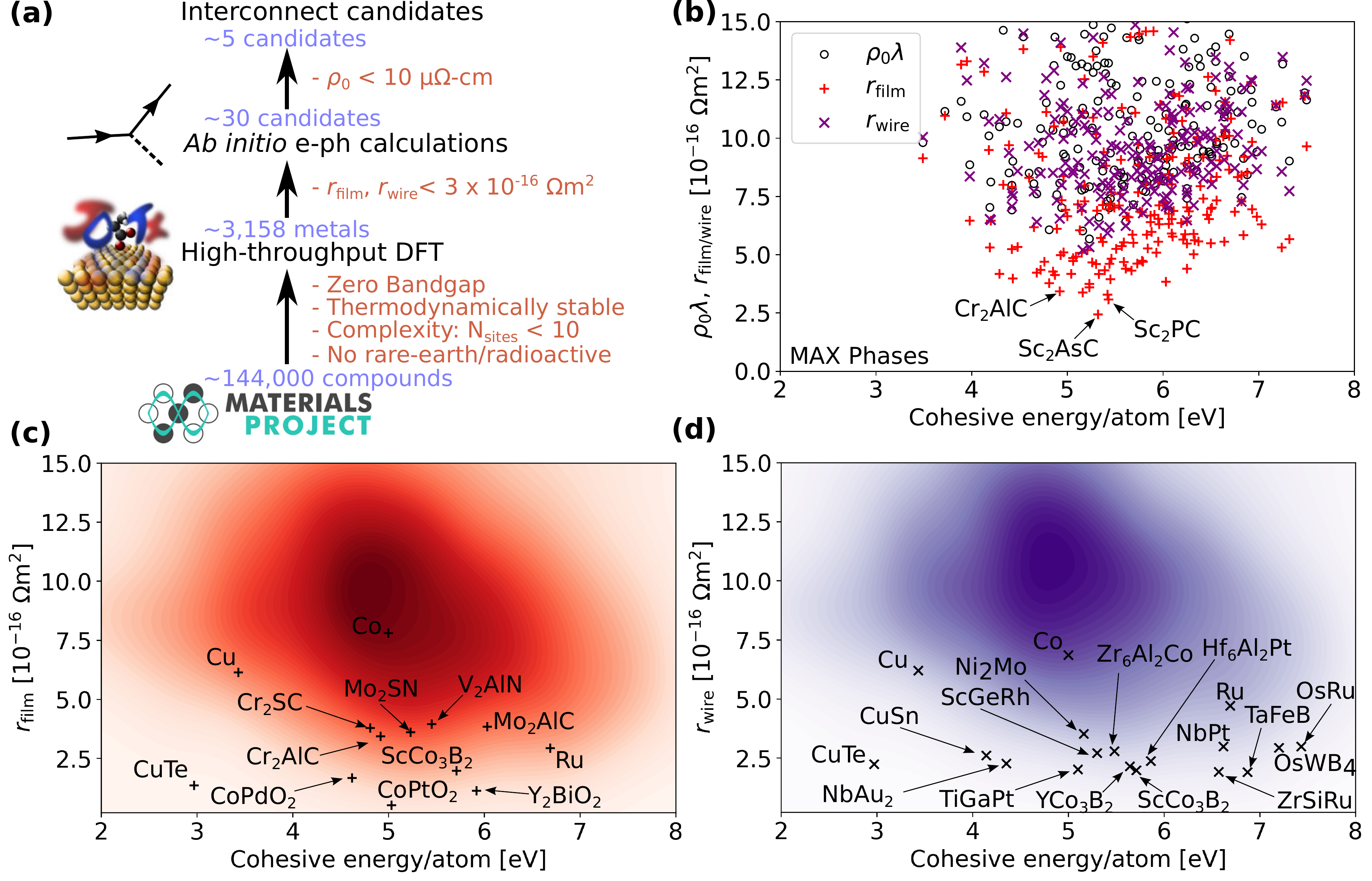}
\caption{(a) We filter stable earth-abundant metals using first-principles calculations of the resistivity scaling coefficients $r\sub{film}$ and $r\sub{wire}$ derived here (Eq.~\ref{eqn:r_film} and \ref{eqn:r_wire}), and evaluate bulk resistivity $\rho_0$ for short-listed candidates using electron-phonon scattering calculations.
(b) Anisotropic MAX phase conductors exhibit much lower resistivity increase in films given by $r\sub{film}$, than predicted by $\rho_0\lambda$, but lose this advantage in wires (higher $r\sub{wire}$).
Most metals exhibit higher (c) $r\sub{film}$ and (d) $r\sub{wire}$ than Cu as shown by the probability density (using kernel density estimation) of the 3106 calculated values, but several intermetallic, oxide and boride candidates (labeled) are lower than Cu.
Cohesive energy on the $x$-axis of (b-d) serves as a proxy for stability against electromigration (higher is better).}
\label{fig:fom}
\end{figure*}

\textbf{High-throughput Screening:}
We use $r\sub{film}$ and $r\sub{wire}$ to search for the most conductive materials in thin films and narrow wires, starting from the Materials Project database with computed structures and electronic properties for over $\sim 144,000$ compounds.\cite{Jain2013}
As indicated in Figure~\ref{fig:fom}(a), we first filter materials of interest based on properties already computed in this database.
Zero band gap filters the number to $\sim 66,000$ metals.
Restricting to thermodynamically stable materials (energy within 0.02 eV of the convex hull to accommodate for DFT errors), excluding rare earth / radioactive elements, and focusing on materials with $< 10$ atoms per primitive unit cell (more complex unit cells are less likely to be reliably synthesized) brings this number down to 3106 candidates.
We also include 214 MAX phase structures that are known for their layered structures and anisotropic electronic properties.\cite{barsoum2011elastic, higashi2018anisotropic, haddad2008dielectric, yao2020exploration, khadzhai2018electrical, ouisse2017magnetotransport, ito2017electronic}
For all these 3320 structures, we perform DFT calculations of $\rho_0\lambda$, $r\sub{film}$ and $r\sub{wire}$ as detailed in the Methods section, and include a list of all calculated materials and properties in the SI.

Figure \ref{fig:fom}(b) compares the predictions of $\rho_0\lambda$, $r\sub{film}$ and $r\sub{wire}$ for all the MAX phase structures, plotted against the cohesive energy per atom, which measures the stability of the material and serves as a proxy for resistance against electromigration (higher is better).
For all of these materials, the in-plane direction is strongly preferred for transport with $v_x, v_y \gg v_z$.
Therefore, with $\hat{j}$ in the $xy$-plane and $\hat{n}$ along $z$, the velocity components along $\hat{n}$ are very small.
This leads to $r\sub{film} \ll \rho_0\lambda$ as discussed above for \CrAlC above.
However, for the case of wires, we can only make the component of velocity along one of $\hat{n}_1$ and $\hat{n}_2$ small -- whichever is along $z$ -- and the other component in-plane remains larger.
Consequently, $r\sub{wire} \gg r\sub{film}$ and becomes comparable to $\rho_0\lambda$.
Therefore, many of the MAX phases are expected to be excellent conductors in thin film geometries, but not in narrow wires necessary for semiconductor interconnects.

Figures~\ref{fig:fom}(c) and (d) respectively plot $r\sub{film}$ and $r\sub{wire}$ against cohesive energy for all the metals shortlisted from Materials Project.
They display a selection of the most promising candidates (low $r$ values) along with the probability distribution of the 3106 calculated points, calculated using kernel density estimation.
Most calculated materials have both $r\sub{film}$ and $r\sub{wire}$ larger than copper, we find $\sim 30$ promising candidates with low values for at least one of $r\sub{film}$ or $r\sub{wire}$.
(In contrast, no material exhibits a lower $\rho_0\lambda$ than the best elemental metals.)
The promising candidates also span different material classes, including intermetallics (CoSn, OsRu, \CrNi, \MoNi and CuPt), borides (\ScCoB, \YCoB and Mn$_2$B) and oxides (\PdCoO and \PtCoO).
Notably, two oxides with the delafossite structure -- \PdCoO and \PtCoO exhibit $r\sub{film} = 1.66$\rlUnit and 0.50\rlUnit respectively, the latter of which is 10$\times$ lower than that of Cu.
For square wires, we find 30 metals with $r\sub{wire}$ less than 3\rlUnit ($\sim$ half that of copper) and 5 metals with $r\sub{wire}$ less than 2\rlUnit ($\sim$ a third that of copper).
(See Table S1 and SI for complete list of calculated properties.)

\begin{figure*}[t]
\centering
\includegraphics[width=\textwidth]{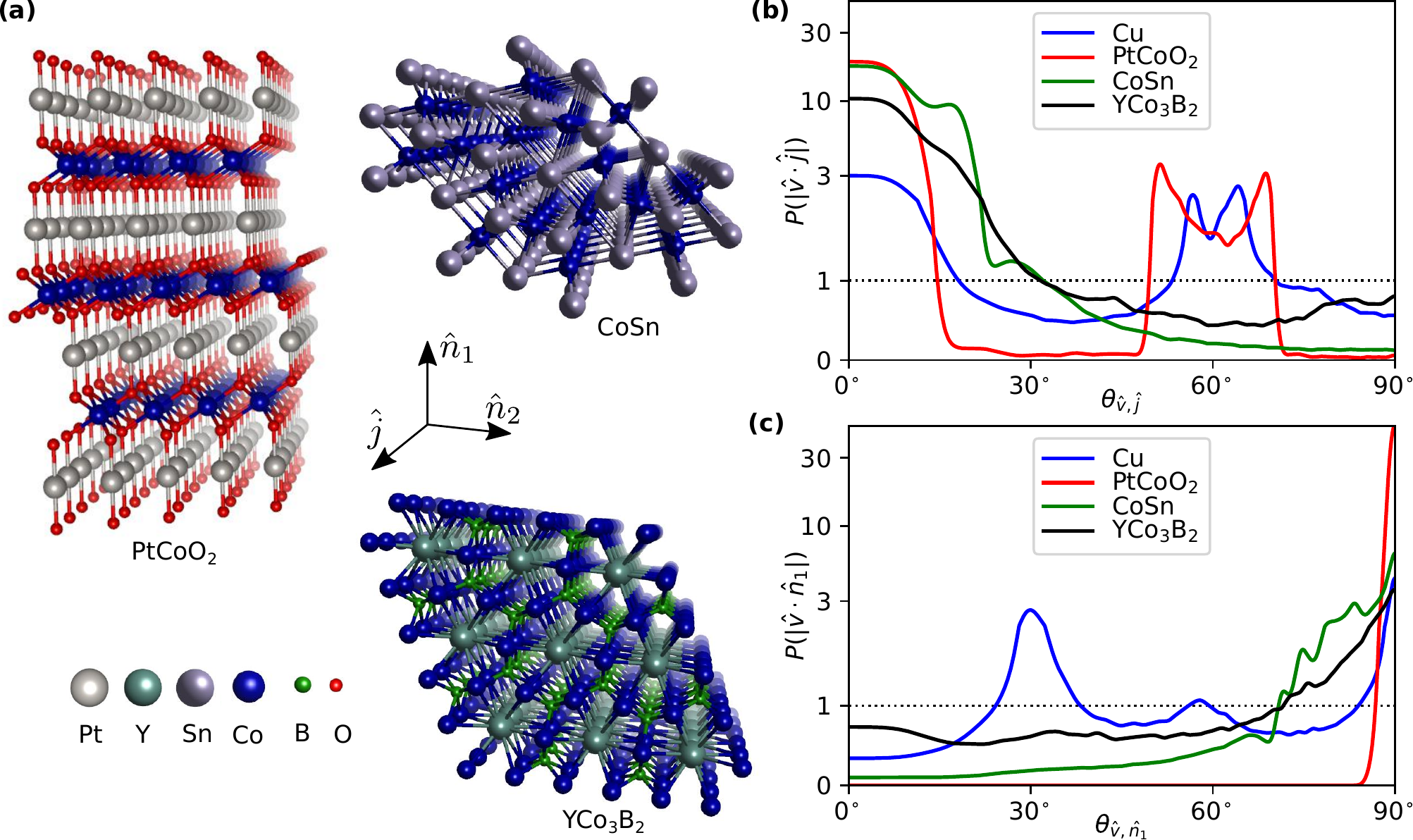}
\caption{(a) Crystal structures of the best candidates for low-resistivity thin films -- \PtCoO and square wires -- CoSn and \YCoB, aligned to show the best transport direction $\hat{j}$, surface normal $\hat{n}_1$ and side walls $\hat{n}_2$ (for wires).
(b) The anisotropic conductors have much higher concentration of velocities near the transport direction, \textit{i.e.} near $\theta_{\hat{v}\,\hat{j}} = 0^\circ$, compared to Cu.
(c) Correspondingly, the anisotropic conductors have higher concentration of velocities perpendicular to the surface normal, \textit{i.e.} near $\theta_{\hat{v}\,\hat{n}_1} = 90^\circ$.
This leads to lower $r\sub{film}$ and $r\sub{wire}$ values, and hence lower expected resistivities at nanoscale dimensions.}
\label{fig:structure}
\end{figure*}

We next examine the connection between the resistivity scaling coefficients, structure and Fermi surfaces of materials, taking as examples the best film candidate, \PtCoO, and the two best wire candidates, CoSn and \YCoB, identified by our first-principles search above.
Figure \ref{fig:structure}(a) shows the crystal structure of these three materials, oriented to indicate the best current direction and corresponding surface / side-wall normal directions.
All three materials are hexagonal, but \PtCoO is layered with the best transport direction along the sheets of Pt atoms \cite{eyert2008metallic,ong2010origin,ong2010unusual}, while CoSn and \YCoB have the best transport direction down the line of Co atoms in each\cite{meier2020flat}.

Figures~\ref{fig:structure}(b) and (c) compare the distribution of velocity directions relative to the transport ($\hat{j}$) and normal ($\hat{n}_1$) directions respectively on the Fermi surface of each of these anisotropic conductors, compared to Cu.
A perfectly spherical Fermi surface would lead to probability density, $P(\cos\theta) = 1$, for angle $\theta$ measured to any axis.
Cu is closest to this limit, while the anisotropic conductors focus their velocities along the transport direction ($\theta_{\vec{v},\hat{j}} = 0^\circ$) and perpendicular to the surface normal direction ($\theta_{\vec{v},\hat{n}_1} = 90^\circ$).
For \PtCoO, the velocities are restricted to a very small range of $\theta_{\vec{v},\hat{n}_1}$ near $90^\circ$ (Figures~\ref{fig:structure}(c)), which leads to $r\sub{film} \ll \rho_0\lambda$.
However, the velocity angle distribution from the transport direction, $\theta_{\vec{v},\hat{j}}$, peaks at both $0^\circ$ and $60^\circ$ in Figures~\ref{fig:structure}(b) due to the in-plane hexagonal symmetry.
This leads to significant side-wall scattering and a high $r\sub{wire}$, analogous to the MAX phases discussed above.
In comparison, the favored wire candidates, CoSn and \YCoB have the velocity direction distributions centered on $\theta_{\vec{v},\hat{j}} = 0^\circ$ as well as $\theta_{\vec{v},\hat{n}_1} = 90^\circ$, ensuring a low value of $r\sub{wire}$.

\begin{figure}[t!]
\includegraphics[width=\columnwidth]{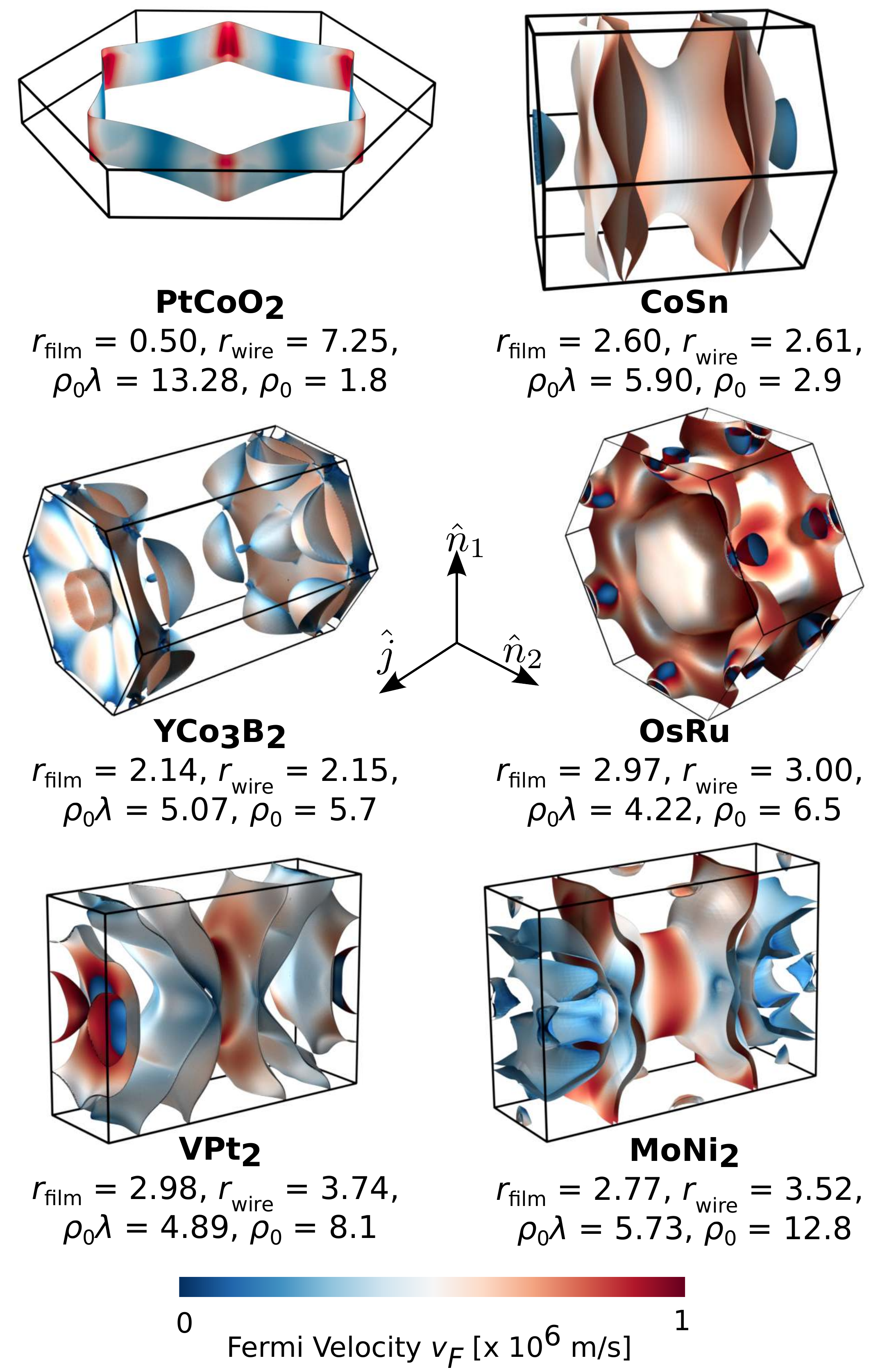}
\caption{Fermi surfaces of the best nanoscale conductors, shortlisted by resistivity scaling coefficients $r\sub{film}$ and $r\sub{wire}$ (in $10^{-16}~\Omega$m$^2$), and then filtered by bulk resistivity $\rho_0$ (in $\mu\Omega\cdot$cm) computed from first-principles.
(Color indicates Fermi velocity magnitude.)
All tend to have flat surfaces perpendicular to the best direction of current flow, leading to large Fermi velocities directed along the transport direction $\hat{j}$ and much smaller velocity components along the surface normal $\hat{n}_1$ and side wall $\hat{n}_2$ directions (See Figure~S1 for Fermi surfaces of additional candidates).}
\label{fig:fermi_surfaces} 
\end{figure}

Figure~\ref{fig:fermi_surfaces} shows the Fermi surfaces of the above anisotropic conductors along with a few others with the lowest $r\sub{wire}$ values (and that remain promising after accounting for bulk resistivity $\rho_0$ discussed next).
In all cases, the Fermi surfaces exhibit almost flat surfaces perpendicular to the direction of current flow $\hat{j}$, indicating that the Fermi velocity is along $\hat{j}$ as shown previously in Figure~\ref{fig:structure}(b).
The \PtCoO Fermi surface in particular is almost perfectly a hexagonal prism, with velocities almost perfectly in the $xy$-plane that leads to the minuscule $r\sub{film}$ (but unremarkable $r\sub{wire}$).
The remaining Fermi surfaces in Figure~\ref{fig:fermi_surfaces} for the wire candidates all exhibit multiple sheets normal to the $\hat{j}$ direction, leading to the low $r\sub{wire}$ that makes them promising for low-resistance wires.

\begin{figure}[t!]
\centering
\includegraphics[width=3in]{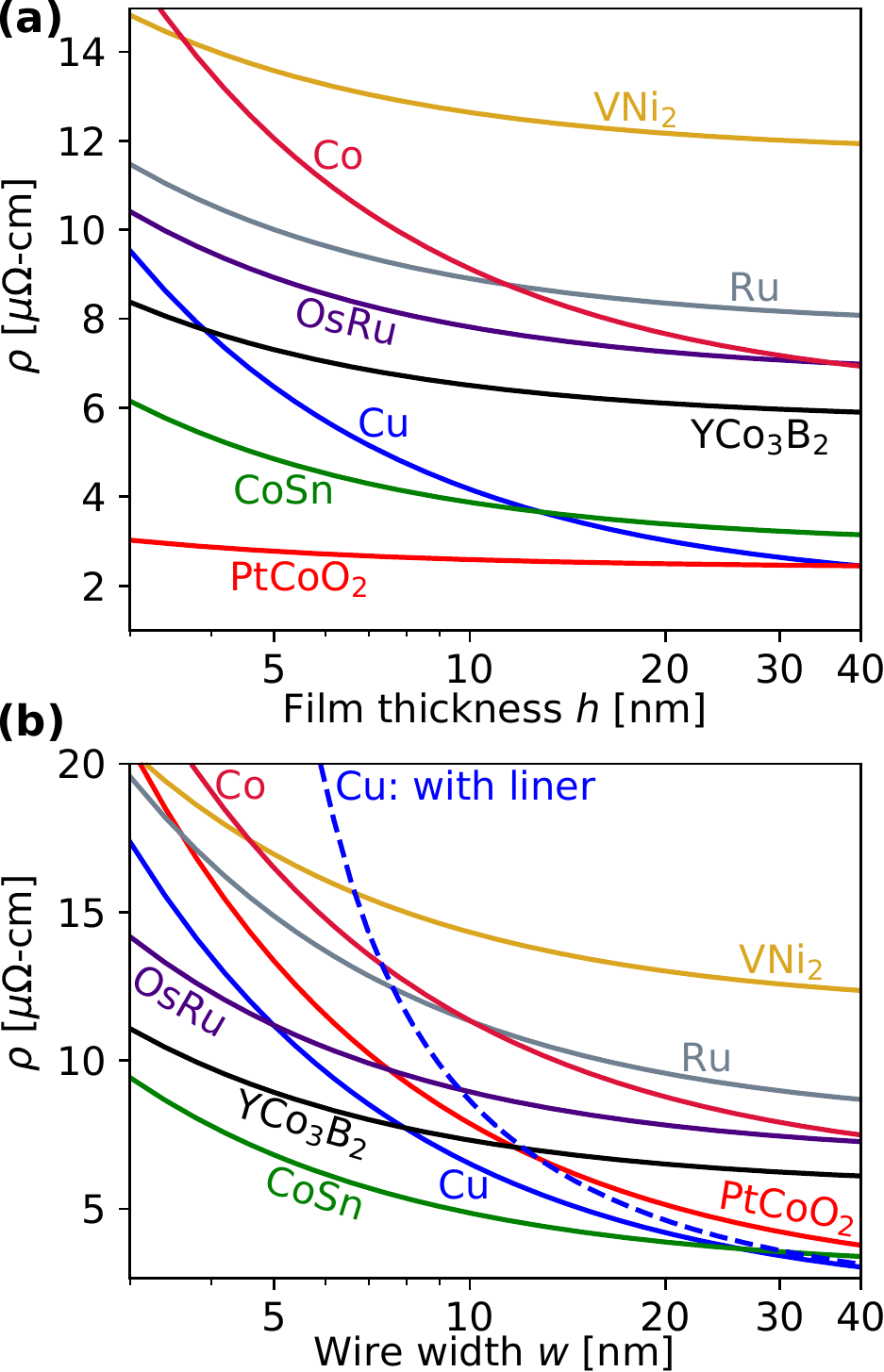}
\caption{Predicted electrical resistivity as a function of (a) film thickness and (b) square wire width for the best nanoscale conductors.
\PtCoO is expected to outperform Cu for films thinner than 40~nm, while CoSn is expected to surpass Cu for films thinner than 13~nm as well as square wires narrower than 25~nm.}
\label{fig:rho}
\end{figure}

\vspace{0.1in}\textbf{Final selection by overall resistivity:}
The high-throughput screening so far using $r\sub{film}$ and $r\sub{wire}$ capture the increase of resistivity with reducing dimensions (in the second term of the F-S model, Eq.~\ref{eqn:FS}).
A promising interconnect material should additionally exhibit low enough bulk resistivity $\rho_0$ that the overall resistance at some film or wire dimension is competitive compared to Cu.
Consequently, the final stage in our computational screening (Figure~\ref{fig:fom}(a)) is the prediction of $\rho_0$ using first-principles electron-phonon scattering calculations (see Methods).
Figure~\ref{fig:rho} shows the overall dimension-dependent resistivity of films and square wires for the most promising materials using the F-S model with both $\rho_0$ and $r\sub{film}$ or $r\sub{wire}$ calculated fully from first-principles, while Table~\ref{tab:results} summarizes the calculated parameters for these candidates.

\begin{table*}[t!]
\begin{center}
\begin{tabular}{ |c|c|c|c|c|c|c|c|c|c| } 
\hline
\multirow{2}{*}{Material} & $v_F$ & $\lambda$ &  $\rho_0 $ & $\rho_0\lambda$  & $r\sub{film}$ & $r\sub{wire}$  & Cohesive energy \\
\cline{5-7}
& [$10^{6}$ m/s] & [nm] &  [$\mu\Omega\cdot$cm] & \multicolumn{3}{c|}{[\rlUnit]}  & [eV/atom]  \\
\hline
Cu & 1.2 & 34.8 & 1.8 & 6.7  & 6.1 & 6.2 & 3.4  \\
\CrAlC & 0.3 & 5.6 & 14.5 & 8.0 & 3.4 & 6.5 & 4.9  \\
IrRu & 0.7 & 6.2 & 8.3 & 5.1 & 3.3 & 5.1 & 8.4\\
CuPt & 0.8 & 11.2 & 6.1 & 6.8 & 3.3 & 5.1 & 4.6 \\
\NiIr & 0.4 & 3.4 & 10.1 & 4.8 &3.2 &3.5 & 6.5 \\
\VPt & 0.5 & 5.4 & 8.1 & 4.9& 3.0 & 3.8 & 5.8 \\
IrRh & 0.7 & 5.8 & 6.6 & 3.6 & 3.0& 3.5 & 6.4\\
\hlcell OsRu & 0.7 & 6.5 & \hlcell 6.5 & 4.2 & \hlcell 3.0 & \hlcell 3.0 & 7.4 \\
\MoNi & 0.4 & 5.7 & 12.8 & 5.7 & 2.8 & 3.5 & 5.2\\
\CrNi & 0.3 & 2.9 & 25.9 & 4.8 & 2.7 & 3.4 & 4.4  \\
\hlcell CoSn & 0.6 & 19.6 & \hlcell 2.9 & 5.9 & \hlcell 2.6 & \hlcell 2.6 & 4.4 \\
\VNi & 0.3 & 3.9 & 13.9 & 4.5 & 2.5 & 3.5 & 5.0 \\
\hlcell\YCoB & 0.4 & 7.6 & \hlcell 5.7 & 5.1 & \hlcell 2.1 & \hlcell 2.2 & 5.6\\
\ScCoB & 0.4 & 5.5 & 8.1 & 5.9 & 2.0 & 2.0 & 5.7 \\
\hlcell\PtCoO & 0.9 & 110.4 & \hlcell 1.8 & 13.3 & \hlcell 0.5 & 7.3 & 5.0  \\
\hline
\end{tabular}
\caption{First-principles predictions of Fermi velocity $v_F$, electron mean free path $\lambda$, bulk resistivity $\rho_0$, resistivity scaling coefficients $\rho_0 \lambda$, $r\sub{film}$ and $r\sub{wire}$ and cohesive energy per atom of promising candidates identified here, with the most promising cases among them highlighted.
Note that $\rho_0$ and $\rho_0\lambda$ correspond to the best transport direction.
See SI for complete listing include specification of crystal orientations and other components of $\rho_0$ and $\rho_0\lambda$ tensors.}
\label{tab:results}
\end{center}
\end{table*}

Figure~\ref{fig:rho} underscores the success of the new resistivity scaling coefficients in identifying candidates which show unusually low increase in resistivity with decreasing thickness.
Several shortlisted candidates like \VNi, \YCoB, CoSn and \PtCoO exhibit a noticeable increase in resistivity only below 10 nm, especially for thin films (Figure~\ref{fig:rho}(a)).
Note that while several metals (e.g. \VNi) show a slower resistivity increase than Cu, they lose the advantage because of their larger `baseline' resistivity $\rho_0$ and will only beat Cu at impractically small dimensions.
Of the new candidates with low bulk resistivity $\rho_0$, \PtCoO stands apart with $\rho_0$ and $v_F$ comparable to Cu, while its $r\sub{film}$ is an order of magnitude smaller, making it the ideal material for thin films.
However, as discussed above, it loses the advantage for wires due to side wall scattering.

For square wires relevant for interconnects, CoSn, \YCoB, \ScCoB and OsRu promise to outperform Cu.
Additionally, all these candidates have a cohesive energy greater than Cu, indicating the possibility that these materials are more stable against electromigration and could be usable without liners.
Further, the ceramic candidates including the oxides and borides may exhibit better stability and surface properties than indicated by the cohesive energy alone.
Figure~\ref{fig:rho}(b) shows that the additional potential advantage in effective resistivity of these materials compared to Cu with a liner, if the new materials can be used without it.
In particular, we expect CoSn to outperform Cu with a liner by 2$\times$ for 10 nm wires, and by 4$\times$ for 6~nm wires.

\textbf{Conclusions:}
We have derived new descriptors of the increased resistivity of metals at nanoscale dimensions that account for anisotropy and directionality of Fermi velocities.
From first-principles evaluation of these descriptors for thousands of materials, we have identified new materials that can exploit this velocity directionality to potentially outperform copper significantly in interconnects for future semiconductor devices.
In particular, we find \PtCoO, which has recently attracted significant attention as a material with potential for hydrodynamic transport at low temperatures \cite{moll2016evidence}, to be the strongest candidate for thin films.
For wires, intermetallics including CoSn and OsRu, as well as borides including \YCoB and \ScCoB are the most promising materials with velocities concentrated near a single transport axis, allowing them to simultaneously minimize scattering from top/bottom surfaces and side walls.
Experimental validation of the predicted superior resistivity in nanoscale single-crystal films and wires, as well as computational understanding of the impact of defects and grain boundaries on the resistivity of directional conductors is necessary to realize the potential of these materials for future interconnects.

\textbf{Methods:}
We use the open-source plane-wave density-functional theory, JDFTx,\cite{JDFTx} to perform DFT calculations with the Perdew-Burke-Ernzerhof exchange-correlation functional,\cite{perdew1996generalized} using non-relativistic ultrasoft pseudopotentials at 20 and 100 Hartree wavefunction and charge density cutoffs,\cite{garrity2014pseudopotentials} of each material with structure obtained from Materials Project.\cite{Jain2013}
We use a $\vec{k}$-point sampling selected automatically such that the effective length of $\vec{k}$-sampled supercell exceeds 100~\AA~in each dimension, and use a Fermi smearing of 0.1~eV.

We perform spin-polarized calculations for metals whose magnetic moments have been reported to be greater than of 0.05 $\mu_\mathrm{B}$ in the Materials Project database.
We do not include spin-orbit coupling as it would be computationally prohibitive for this high-throughput search, and in any case, most candidates do not include heavy elements for which the effects of spin-orbit coupling would become pronounced.

With electronic energies $\varepsilon_{\vec{k}b}$ and corresponding velocities $\vec{v}_{\vec{k}b}$ computed from the expectation value of commutator $-i[\vec{r},\hat{H}]/\hbar$ to account for nonlocal pseudopotential contributions,\cite{brown2016nonradiative} we directly evaluate the Brillouin zone integrals in Eqs.~\ref{eqn:rho_lambda}, \ref{eqn:g1} and \ref{eqn:g2} on the fine DFT $k$-mesh.
For $r\sub{film}$ and $r\sub{wire}$, we additionally optimize a rotation matrix to identify the best combination of $\hat{j}$, $\hat{n}_1$ and $\hat{n}_2$ for each material.
For benchmarking, we compare the values of the $\rho_0 \lambda$ descriptor computed using this methodology with the first-principles calculations of Gall\cite{gall2016electron} and find very good agreement (See Table~S2 in SI).

For shortlisted materials, we perform first-principles electron-phonon scattering calculations to evaluate the bulk resistivity $\rho_0$ using JDFTx.\cite{JDFTx, brown2016ab, habib2018hot}
Briefly, we calculate phonons with a $q$-point mesh such that the effective length of the phonon supercell exceeds 15~\AA~in each dimension, and use maximally localized Wannier functions to interpolate electron, phonon and electron-phonon matrix elements to fine meshes exceeding 100 points in each dimension.
We then compute the electron-phonon momentum relaxation time $\tau_{\vec{k}b}$ of each electronic state, and compute the bulk conductivity using the Boltzmann equation in the per-band relaxation time approximation.
See Ref.~\cite{habib2018hot} for details on the electron-phonon scattering calculation method.

\textbf{Acknowledgements:}
The authors acknowledge funding from SRC under Task No. 2966.
Calculations were carried out at the Center for Computational Innovations at Rensselaer Polytechnic Institute.

\providecommand{\latin}[1]{#1}
\makeatletter
\providecommand{\doi}
  {\begingroup\let\do\@makeother\dospecials
  \catcode`\{=1 \catcode`\}=2 \doi@aux}
\providecommand{\doi@aux}[1]{\endgroup\texttt{#1}}
\makeatother
\providecommand*\mcitethebibliography{\thebibliography}
\csname @ifundefined\endcsname{endmcitethebibliography}
  {\let\endmcitethebibliography\endthebibliography}{}


\begin{mcitethebibliography}{47}
\providecommand*\natexlab[1]{#1}
\providecommand*\mciteSetBstSublistMode[1]{}
\providecommand*\mciteSetBstMaxWidthForm[2]{}
\providecommand*\mciteBstWouldAddEndPuncttrue
  {\def\EndOfBibitem{\unskip.}}
\providecommand*\mciteBstWouldAddEndPunctfalse
  {\let\EndOfBibitem\relax}
\providecommand*\mciteSetBstMidEndSepPunct[3]{}
\providecommand*\mciteSetBstSublistLabelBeginEnd[3]{}
\providecommand*\EndOfBibitem{}
\mciteSetBstSublistMode{f}
\mciteSetBstMaxWidthForm{subitem}{(\alph{mcitesubitemcount})}
\mciteSetBstSublistLabelBeginEnd
  {\mcitemaxwidthsubitemform\space}
  {\relax}
  {\relax}

\bibitem[Salahuddin \latin{et~al.}(2018)Salahuddin, Ni, and
  Datta]{salahuddin2018era}
Salahuddin,~S.; Ni,~K.; Datta,~S. The Era of Hyper-scaling in Electronics.
  \emph{Nature Elec.} \textbf{2018}, \emph{1}, 442--450\relax
\mciteBstWouldAddEndPuncttrue
\mciteSetBstMidEndSepPunct{\mcitedefaultmidpunct}
{\mcitedefaultendpunct}{\mcitedefaultseppunct}\relax
\EndOfBibitem
\bibitem[Ferry(2008)]{ferry2008nanowires}
Ferry,~D.~K. Nanowires in Nanoelectronics. \emph{Science} \textbf{2008},
  \emph{319}, 579--580\relax
\mciteBstWouldAddEndPuncttrue
\mciteSetBstMidEndSepPunct{\mcitedefaultmidpunct}
{\mcitedefaultendpunct}{\mcitedefaultseppunct}\relax
\EndOfBibitem
\bibitem[Banerjee and Mehrotra(2001)Banerjee, and Mehrotra]{banerjee2001global}
Banerjee,~K.; Mehrotra,~A. Global (Interconnect) Warming. \emph{IEEE Circ.
  Dev.} \textbf{2001}, \emph{17}, 16--32\relax
\mciteBstWouldAddEndPuncttrue
\mciteSetBstMidEndSepPunct{\mcitedefaultmidpunct}
{\mcitedefaultendpunct}{\mcitedefaultseppunct}\relax
\EndOfBibitem
\bibitem[Charles~Jr(2005)]{charles2005miniaturized}
Charles~Jr,~H.~K. Miniaturized Electronics. \emph{Johns Hopkins APL Tech. Dig.}
  \textbf{2005}, \emph{26}, 402--413\relax
\mciteBstWouldAddEndPuncttrue
\mciteSetBstMidEndSepPunct{\mcitedefaultmidpunct}
{\mcitedefaultendpunct}{\mcitedefaultseppunct}\relax
\EndOfBibitem
\bibitem[Gall(2020)]{gall2020search}
Gall,~D. The Search for the Most Conductive Metal for Narrow Interconnect
  Lines. \emph{J. Appl. Phys.} \textbf{2020}, \emph{127}, 050901\relax
\mciteBstWouldAddEndPuncttrue
\mciteSetBstMidEndSepPunct{\mcitedefaultmidpunct}
{\mcitedefaultendpunct}{\mcitedefaultseppunct}\relax
\EndOfBibitem
\bibitem[Gall \latin{et~al.}(2021)Gall, Cha, Chen, Han, Hinkle, Robinson,
  Sundararaman, and Torsi]{gall2021bulletin}
Gall,~D.; Cha,~J.~J.; Chen,~Z.; Han,~H.-J.; Hinkle,~C.; Robinson,~J.~A.;
  Sundararaman,~R.; Torsi,~R. Materials for Interconnects. \emph{MRS Bullet.}
  \textbf{2021}, \emph{46}, 959--966\relax
\mciteBstWouldAddEndPuncttrue
\mciteSetBstMidEndSepPunct{\mcitedefaultmidpunct}
{\mcitedefaultendpunct}{\mcitedefaultseppunct}\relax
\EndOfBibitem
\bibitem[Xu \latin{et~al.}(2015)Xu, Wang, Guo, Chen, Liu, and
  Huang]{xu2015copper}
Xu,~W.-H.; Wang,~L.; Guo,~Z.; Chen,~X.; Liu,~J.; Huang,~X.-J. Copper Nanowires
  as Nanoscale Interconnects: Their Stability, Electrical Transport, and
  Mechanical Properties. \emph{ACS Nano} \textbf{2015}, \emph{9},
  241--250\relax
\mciteBstWouldAddEndPuncttrue
\mciteSetBstMidEndSepPunct{\mcitedefaultmidpunct}
{\mcitedefaultendpunct}{\mcitedefaultseppunct}\relax
\EndOfBibitem
\bibitem[Simbeck \latin{et~al.}(2012)Simbeck, Lanzillo, Kharche, Verstraete,
  and Nayak]{simbeck2012aluminum}
Simbeck,~A.~J.; Lanzillo,~N.; Kharche,~N.; Verstraete,~M.~J.; Nayak,~S.~K.
  Aluminum Conducts Better than Copper at the Atomic Scale: A First-principles
  Study of Metallic Atomic Wires. \emph{ACS Nano} \textbf{2012}, \emph{6},
  10449--10455\relax
\mciteBstWouldAddEndPuncttrue
\mciteSetBstMidEndSepPunct{\mcitedefaultmidpunct}
{\mcitedefaultendpunct}{\mcitedefaultseppunct}\relax
\EndOfBibitem
\bibitem[Lanzillo \latin{et~al.}(2014)Lanzillo, Thomas, Watson, Washington, and
  Nayak]{lanzillo2014pressure}
Lanzillo,~N.~A.; Thomas,~J.~B.; Watson,~B.; Washington,~M.; Nayak,~S.~K.
  Pressure-enabled Phonon Engineering in Metals. \emph{Proc. Nat. Acad. Sci.}
  \textbf{2014}, \emph{111}, 8712--8716\relax
\mciteBstWouldAddEndPuncttrue
\mciteSetBstMidEndSepPunct{\mcitedefaultmidpunct}
{\mcitedefaultendpunct}{\mcitedefaultseppunct}\relax
\EndOfBibitem
\bibitem[Lanzillo(2017)]{lanzillo2017ab}
Lanzillo,~N.~A. Ab Initio Evaluation of Electron Transport Properties of Pt,
  Rh, Ir, and Pd Nanowires for Advanced Interconnect Applications. \emph{J.
  Appl. Phys.} \textbf{2017}, \emph{121}, 175104\relax
\mciteBstWouldAddEndPuncttrue
\mciteSetBstMidEndSepPunct{\mcitedefaultmidpunct}
{\mcitedefaultendpunct}{\mcitedefaultseppunct}\relax
\EndOfBibitem
\bibitem[Jiang \latin{et~al.}(2017)Jiang, Kang, Cao, Xie, Zhang, Chu, Liu, and
  Banerjee]{jiang2017intercalation}
Jiang,~J.; Kang,~J.; Cao,~W.; Xie,~X.; Zhang,~H.; Chu,~J.~H.; Liu,~W.;
  Banerjee,~K. Intercalation Doped Multilayer-graphene-nanoribbons for
  Next-generation Interconnects. \emph{Nano Lett.} \textbf{2017}, \emph{17},
  1482--1488\relax
\mciteBstWouldAddEndPuncttrue
\mciteSetBstMidEndSepPunct{\mcitedefaultmidpunct}
{\mcitedefaultendpunct}{\mcitedefaultseppunct}\relax
\EndOfBibitem
\bibitem[Hu \latin{et~al.}(2022)Hu, Conlin, Lee, Kim, and Cho]{hu2022van}
Hu,~Y.; Conlin,~P.; Lee,~Y.; Kim,~D.; Cho,~K. Van der Waals 2D Metallic
  Materials for Low-Resistivity Interconnects. \emph{J. Mater. Chem. C}
  \textbf{2022}, \relax
\mciteBstWouldAddEndPunctfalse
\mciteSetBstMidEndSepPunct{\mcitedefaultmidpunct}
{}{\mcitedefaultseppunct}\relax
\EndOfBibitem
\bibitem[Zhang \latin{et~al.}(2019)Zhang, Ni, Zhang, Yuan, Liu, Zou, Liao, Du,
  Narayan, Zhang, \latin{et~al.} others]{zhang2019ultrahigh}
Zhang,~C.; Ni,~Z.; Zhang,~J.; Yuan,~X.; Liu,~Y.; Zou,~Y.; Liao,~Z.; Du,~Y.;
  Narayan,~A.; Zhang,~H., \latin{et~al.}  Ultrahigh Conductivity in Weyl
  Semimetal NbAs Nanobelts. \emph{Nature Mater.} \textbf{2019}, \emph{18},
  482--488\relax
\mciteBstWouldAddEndPuncttrue
\mciteSetBstMidEndSepPunct{\mcitedefaultmidpunct}
{\mcitedefaultendpunct}{\mcitedefaultseppunct}\relax
\EndOfBibitem
\bibitem[Chen \latin{et~al.}(2020)Chen, Bajpai, Lanzillo, Hsu, Lin, and
  Liang]{chen2020topological}
Chen,~C.-T.; Bajpai,~U.; Lanzillo,~N.~A.; Hsu,~C.-H.; Lin,~H.; Liang,~G.
  Topological Semimetals for Scaled Back-End-Of-Line Interconnect Beyond Cu.
  2020 IEEE Int. Elec. Dev. Meet. (IEDM). 2020; pp 32--4\relax
\mciteBstWouldAddEndPuncttrue
\mciteSetBstMidEndSepPunct{\mcitedefaultmidpunct}
{\mcitedefaultendpunct}{\mcitedefaultseppunct}\relax
\EndOfBibitem
\bibitem[Han \latin{et~al.}(2021)Han, Liu, and Cha]{han20211d}
Han,~H.~J.; Liu,~P.; Cha,~J.~J. 1D Topological Systems for Next-generation
  Electronics. \emph{Matter} \textbf{2021}, \emph{4}, 2596--2598\relax
\mciteBstWouldAddEndPuncttrue
\mciteSetBstMidEndSepPunct{\mcitedefaultmidpunct}
{\mcitedefaultendpunct}{\mcitedefaultseppunct}\relax
\EndOfBibitem
\bibitem[Philip \latin{et~al.}(2016)Philip, Hirsbrunner, Park, and
  Gilbert]{philip2016performance}
Philip,~T.~M.; Hirsbrunner,~M.~R.; Park,~M.~J.; Gilbert,~M.~J. Performance of
  Topological Insulator Interconnects. \emph{IEEE Elec. Dev. Lett.}
  \textbf{2016}, \emph{38}, 138--141\relax
\mciteBstWouldAddEndPuncttrue
\mciteSetBstMidEndSepPunct{\mcitedefaultmidpunct}
{\mcitedefaultendpunct}{\mcitedefaultseppunct}\relax
\EndOfBibitem
\bibitem[Fuchs(1938)]{fuchs1938math}
Fuchs,~K. The Conductivity of Thin Metallic Films According to the Electron
  Theory of Metals. \emph{Math. Proc. Cambridge Philos. Soc.} \textbf{1938},
  \emph{34}, 100\relax
\mciteBstWouldAddEndPuncttrue
\mciteSetBstMidEndSepPunct{\mcitedefaultmidpunct}
{\mcitedefaultendpunct}{\mcitedefaultseppunct}\relax
\EndOfBibitem
\bibitem[Sondheimer(1952)]{sondheimer1952adv}
Sondheimer,~E. The Mean Free Path of Electrons in Metals. \emph{Adv. Phys.}
  \textbf{1952}, \emph{1}, 1--42\relax
\mciteBstWouldAddEndPuncttrue
\mciteSetBstMidEndSepPunct{\mcitedefaultmidpunct}
{\mcitedefaultendpunct}{\mcitedefaultseppunct}\relax
\EndOfBibitem
\bibitem[Mayadas and Shatzkes(1970)Mayadas, and
  Shatzkes]{mayadas1970electrical}
Mayadas,~A.; Shatzkes,~M. Electrical-resistivity Model for Polycrystalline
  Films: the Case of Arbitrary Reflection at External Surfaces. \emph{Phys.
  Rev. B} \textbf{1970}, \emph{1}, 1382\relax
\mciteBstWouldAddEndPuncttrue
\mciteSetBstMidEndSepPunct{\mcitedefaultmidpunct}
{\mcitedefaultendpunct}{\mcitedefaultseppunct}\relax
\EndOfBibitem
\bibitem[Gall(2016)]{gall2016electron}
Gall,~D. Electron Mean Free Path in Elemental Metals. \emph{J. Appl. Phys.}
  \textbf{2016}, \emph{119}, 085101\relax
\mciteBstWouldAddEndPuncttrue
\mciteSetBstMidEndSepPunct{\mcitedefaultmidpunct}
{\mcitedefaultendpunct}{\mcitedefaultseppunct}\relax
\EndOfBibitem
\bibitem[Chen \latin{et~al.}(2018)Chen, Ando, Sutou, Gall, and
  Koike]{chen2018nial}
Chen,~L.; Ando,~D.; Sutou,~Y.; Gall,~D.; Koike,~J. NiAl as a Potential Material
  for Liner-and Barrier-free Interconnect in Ultrasmall Technology Node.
  \emph{Appl. Phys. Lett.} \textbf{2018}, \emph{113}, 183503\relax
\mciteBstWouldAddEndPuncttrue
\mciteSetBstMidEndSepPunct{\mcitedefaultmidpunct}
{\mcitedefaultendpunct}{\mcitedefaultseppunct}\relax
\EndOfBibitem
\bibitem[Chen \latin{et~al.}(2021)Chen, Kumar, Yahagi, Ando, Sutou, Gall,
  Sundararaman, and Koike]{chen2021intermetallics}
Chen,~L.; Kumar,~S.; Yahagi,~M.; Ando,~D.; Sutou,~Y.; Gall,~D.;
  Sundararaman,~R.; Koike,~J. Interdiffusion Reliability and Resistivity
  Scaling of Intermetallic Compounds as Advanced Interconnect Materials.
  \emph{J. Appl. Phys.} \textbf{2021}, \emph{129}, 035301\relax
\mciteBstWouldAddEndPuncttrue
\mciteSetBstMidEndSepPunct{\mcitedefaultmidpunct}
{\mcitedefaultendpunct}{\mcitedefaultseppunct}\relax
\EndOfBibitem
\bibitem[Kaloyeros and Eisenbraun(2000)Kaloyeros, and
  Eisenbraun]{kaloyeros2000ultrathin}
Kaloyeros,~A.; Eisenbraun,~E. Ultrathin Diffusion Barriers/Liners for Gigascale
  Copper Metallization. \emph{Ann. Rev. Mater. Sci.} \textbf{2000}, \emph{30},
  363--385\relax
\mciteBstWouldAddEndPuncttrue
\mciteSetBstMidEndSepPunct{\mcitedefaultmidpunct}
{\mcitedefaultendpunct}{\mcitedefaultseppunct}\relax
\EndOfBibitem
\bibitem[Sankaran \latin{et~al.}(2021)Sankaran, Moors, T{\H{o}}kei, Adelmann,
  and Pourtois]{sankaran2021ab}
Sankaran,~K.; Moors,~K.; T{\H{o}}kei,~Z.; Adelmann,~C.; Pourtois,~G. Ab Initio
  Screening of Metallic MAX Ceramics for Advanced Interconnect Applications.
  \emph{Phys. Rev. Mater.} \textbf{2021}, \emph{5}, 056002\relax
\mciteBstWouldAddEndPuncttrue
\mciteSetBstMidEndSepPunct{\mcitedefaultmidpunct}
{\mcitedefaultendpunct}{\mcitedefaultseppunct}\relax
\EndOfBibitem
\bibitem[Zhang \latin{et~al.}(2021)Zhang, Kumar, Sundararaman, and
  Gall]{zhang2021resistivity}
Zhang,~M.; Kumar,~S.; Sundararaman,~R.; Gall,~D. Resistivity Scaling in
  Epitaxial MAX-phase Ti4SiC3 (0001) Layers. \emph{J. Appl. Phys.}
  \textbf{2021}, \emph{130}, 034302\relax
\mciteBstWouldAddEndPuncttrue
\mciteSetBstMidEndSepPunct{\mcitedefaultmidpunct}
{\mcitedefaultendpunct}{\mcitedefaultseppunct}\relax
\EndOfBibitem
\bibitem[Brown \latin{et~al.}(2016)Brown, Sundararaman, Narang, Goddard~III,
  and Atwater]{brown2016nonradiative}
Brown,~A.~M.; Sundararaman,~R.; Narang,~P.; Goddard~III,~W.~A.; Atwater,~H.~A.
  Nonradiative Plasmon Decay and Hot Carrier Dynamics: Effects of Phonons,
  Surfaces, and Geometry. \emph{ACS Nano} \textbf{2016}, \emph{10},
  957--966\relax
\mciteBstWouldAddEndPuncttrue
\mciteSetBstMidEndSepPunct{\mcitedefaultmidpunct}
{\mcitedefaultendpunct}{\mcitedefaultseppunct}\relax
\EndOfBibitem
\bibitem[Ito \latin{et~al.}(2017)Ito, Pinek, Fujita, Nakatake, Ideta, Tanaka,
  and Ouisse]{ito2017electronic}
Ito,~T.; Pinek,~D.; Fujita,~T.; Nakatake,~M.; Ideta,~S.-i.; Tanaka,~K.;
  Ouisse,~T. Electronic Structure of Cr$_2$AlC as Observed by Angle-resolved
  Photoemission Spectroscopy. \emph{Phys. Rev. B} \textbf{2017}, \emph{96},
  195168\relax
\mciteBstWouldAddEndPuncttrue
\mciteSetBstMidEndSepPunct{\mcitedefaultmidpunct}
{\mcitedefaultendpunct}{\mcitedefaultseppunct}\relax
\EndOfBibitem
\bibitem[Ouisse and Barsoum(2017)Ouisse, and
  Barsoum]{ouisse2017magnetotransport}
Ouisse,~T.; Barsoum,~M.~W. Magnetotransport in the MAX Phases and their 2D
  Derivatives: MXenes. \emph{Mater. Res. Lett.} \textbf{2017}, \emph{5},
  365--378\relax
\mciteBstWouldAddEndPuncttrue
\mciteSetBstMidEndSepPunct{\mcitedefaultmidpunct}
{\mcitedefaultendpunct}{\mcitedefaultseppunct}\relax
\EndOfBibitem
\bibitem[Lucas(1965)]{lucas1965electrical}
Lucas,~M. Electrical Conductivity of Thin Metallic Films with Unlike Surfaces.
  \emph{J. Appl. Phys.} \textbf{1965}, \emph{36}, 1632--1635\relax
\mciteBstWouldAddEndPuncttrue
\mciteSetBstMidEndSepPunct{\mcitedefaultmidpunct}
{\mcitedefaultendpunct}{\mcitedefaultseppunct}\relax
\EndOfBibitem
\bibitem[Zheng and Gall(2017)Zheng, and Gall]{zheng2017anisotropic}
Zheng,~P.; Gall,~D. The Anisotropic Size Effect of the Electrical Resistivity
  of Metal Thin Films: Tungsten. \emph{J. Appl. Phys.} \textbf{2017},
  \emph{122}, 135301\relax
\mciteBstWouldAddEndPuncttrue
\mciteSetBstMidEndSepPunct{\mcitedefaultmidpunct}
{\mcitedefaultendpunct}{\mcitedefaultseppunct}\relax
\EndOfBibitem
\bibitem[Jain \latin{et~al.}(2013)Jain, Ong, Hautier, Chen, Richards, Dacek,
  Cholia, Gunter, Skinner, Ceder, and Persson]{Jain2013}
Jain,~A.; Ong,~S.~P.; Hautier,~G.; Chen,~W.; Richards,~W.~D.; Dacek,~S.;
  Cholia,~S.; Gunter,~D.; Skinner,~D.; Ceder,~G.; Persson,~K.~a. {The Materials
  Project: A Materials Genome Approach to Accelerating Materials Innovation}.
  \emph{APL Mater.} \textbf{2013}, \emph{1}, 011002\relax
\mciteBstWouldAddEndPuncttrue
\mciteSetBstMidEndSepPunct{\mcitedefaultmidpunct}
{\mcitedefaultendpunct}{\mcitedefaultseppunct}\relax
\EndOfBibitem
\bibitem[Barsoum and Radovic(2011)Barsoum, and Radovic]{barsoum2011elastic}
Barsoum,~M.~W.; Radovic,~M. Elastic and Mechanical Properties of the MAX
  Phases. \emph{Ann. Rev. Mater. Res.} \textbf{2011}, \emph{41}, 195--227\relax
\mciteBstWouldAddEndPuncttrue
\mciteSetBstMidEndSepPunct{\mcitedefaultmidpunct}
{\mcitedefaultendpunct}{\mcitedefaultseppunct}\relax
\EndOfBibitem
\bibitem[Higashi \latin{et~al.}(2018)Higashi, Momono, Kishida, Okamoto, and
  Inui]{higashi2018anisotropic}
Higashi,~M.; Momono,~S.; Kishida,~K.; Okamoto,~N.~L.; Inui,~H. Anisotropic
  Plastic Deformation of Single Crystals of the MAX Phase Compound Ti3SiC2
  Investigated by Micropillar Compression. \emph{Acta Mater.} \textbf{2018},
  \emph{161}, 161--170\relax
\mciteBstWouldAddEndPuncttrue
\mciteSetBstMidEndSepPunct{\mcitedefaultmidpunct}
{\mcitedefaultendpunct}{\mcitedefaultseppunct}\relax
\EndOfBibitem
\bibitem[Haddad \latin{et~al.}(2008)Haddad, Garcia-Caurel, Hultman, Barsoum,
  and Hug]{haddad2008dielectric}
Haddad,~N.; Garcia-Caurel,~E.; Hultman,~L.; Barsoum,~M.~W.; Hug,~G. Dielectric
  Properties of Ti$_2$AlC and Ti$_2$AlN MAX Phases: The Conductivity
  Anisotropy. \emph{J. Appl. Phys.} \textbf{2008}, \emph{104}, 023531\relax
\mciteBstWouldAddEndPuncttrue
\mciteSetBstMidEndSepPunct{\mcitedefaultmidpunct}
{\mcitedefaultendpunct}{\mcitedefaultseppunct}\relax
\EndOfBibitem
\bibitem[Yao \latin{et~al.}(2020)Yao, Qian, Li, Li, Zuo, Xu, and
  Li]{yao2020exploration}
Yao,~P.; Qian,~Y.; Li,~W.; Li,~C.; Zuo,~J.; Xu,~J.; Li,~M. Exploration of
  Dielectric and Microwave Absorption Properties of Quaternary MAX Phase
  Ceramic (Cr$_{2/3}$Ti$_{1/3}$)$_3$AlC$_2$. \emph{Ceram. Int.} \textbf{2020},
  \emph{46}, 22919--22926\relax
\mciteBstWouldAddEndPuncttrue
\mciteSetBstMidEndSepPunct{\mcitedefaultmidpunct}
{\mcitedefaultendpunct}{\mcitedefaultseppunct}\relax
\EndOfBibitem
\bibitem[Khadzhai \latin{et~al.}(2018)Khadzhai, Vovk, Prichna, Gevorkyan,
  Kislitsa, and Solovjov]{khadzhai2018electrical}
Khadzhai,~G.~Y.; Vovk,~R.; Prichna,~T.; Gevorkyan,~E.; Kislitsa,~M.;
  Solovjov,~A. Electrical and Thermal Conductivity of the Ti3AlC2 MAX Phase at
  Low Temperatures. \emph{Low Temp. Phys.} \textbf{2018}, \emph{44},
  451--452\relax
\mciteBstWouldAddEndPuncttrue
\mciteSetBstMidEndSepPunct{\mcitedefaultmidpunct}
{\mcitedefaultendpunct}{\mcitedefaultseppunct}\relax
\EndOfBibitem
\bibitem[Eyert \latin{et~al.}(2008)Eyert, Fr{\'e}sard, and
  Maignan]{eyert2008metallic}
Eyert,~V.; Fr{\'e}sard,~R.; Maignan,~A. On the Metallic Conductivity of the
  Delafossites PdCoO2 and PtCoO2. \emph{Chem. Mater.} \textbf{2008}, \emph{20},
  2370--2373\relax
\mciteBstWouldAddEndPuncttrue
\mciteSetBstMidEndSepPunct{\mcitedefaultmidpunct}
{\mcitedefaultendpunct}{\mcitedefaultseppunct}\relax
\EndOfBibitem
\bibitem[Ong \latin{et~al.}(2010)Ong, Zhang, John, and Wu]{ong2010origin}
Ong,~K.~P.; Zhang,~J.; John,~S.~T.; Wu,~P. Origin of Anisotropy and Metallic
  Behavior in Delafossite PdCoO$_2$. \emph{Phys. Rev. B} \textbf{2010},
  \emph{81}, 115120\relax
\mciteBstWouldAddEndPuncttrue
\mciteSetBstMidEndSepPunct{\mcitedefaultmidpunct}
{\mcitedefaultendpunct}{\mcitedefaultseppunct}\relax
\EndOfBibitem
\bibitem[Ong \latin{et~al.}(2010)Ong, Singh, and Wu]{ong2010unusual}
Ong,~K.~P.; Singh,~D.~J.; Wu,~P. Unusual Transport and Strongly Anisotropic
  Thermopower in PtCoO$_2$ and PdCoO$_2$. \emph{Phys. Rev. Lett.}
  \textbf{2010}, \emph{104}, 176601\relax
\mciteBstWouldAddEndPuncttrue
\mciteSetBstMidEndSepPunct{\mcitedefaultmidpunct}
{\mcitedefaultendpunct}{\mcitedefaultseppunct}\relax
\EndOfBibitem
\bibitem[Meier \latin{et~al.}(2020)Meier, Du, Okamoto, Mohanta, May, McGuire,
  Bridges, Samolyuk, and Sales]{meier2020flat}
Meier,~W.~R.; Du,~M.-H.; Okamoto,~S.; Mohanta,~N.; May,~A.~F.; McGuire,~M.~A.;
  Bridges,~C.~A.; Samolyuk,~G.~D.; Sales,~B.~C. Flat Bands in the CoSn-type
  Compounds. \emph{Phys. Rev. B} \textbf{2020}, \emph{102}, 075148\relax
\mciteBstWouldAddEndPuncttrue
\mciteSetBstMidEndSepPunct{\mcitedefaultmidpunct}
{\mcitedefaultendpunct}{\mcitedefaultseppunct}\relax
\EndOfBibitem
\bibitem[Moll \latin{et~al.}(2016)Moll, Kushwaha, Nandi, Schmidt, and
  Mackenzie]{moll2016evidence}
Moll,~P.~J.; Kushwaha,~P.; Nandi,~N.; Schmidt,~B.; Mackenzie,~A.~P. Evidence
  for Hydrodynamic Electron Flow in PdCoO$_2$. \emph{Science} \textbf{2016},
  \emph{351}, 1061--1064\relax
\mciteBstWouldAddEndPuncttrue
\mciteSetBstMidEndSepPunct{\mcitedefaultmidpunct}
{\mcitedefaultendpunct}{\mcitedefaultseppunct}\relax
\EndOfBibitem
\bibitem[Sundararaman \latin{et~al.}(2017)Sundararaman, Letchworth-Weaver,
  Schwarz, Gunceler, Ozhabes, and Arias]{JDFTx}
Sundararaman,~R.; Letchworth-Weaver,~K.; Schwarz,~K.; Gunceler,~D.;
  Ozhabes,~Y.; Arias,~T.~A. JDFTx: Software for Joint Density-Functional
  Theory. \emph{SoftwareX} \textbf{2017}, \emph{6}, 278 -- 284\relax
\mciteBstWouldAddEndPuncttrue
\mciteSetBstMidEndSepPunct{\mcitedefaultmidpunct}
{\mcitedefaultendpunct}{\mcitedefaultseppunct}\relax
\EndOfBibitem
\bibitem[Perdew \latin{et~al.}(1996)Perdew, Burke, and
  Ernzerhof]{perdew1996generalized}
Perdew,~J.~P.; Burke,~K.; Ernzerhof,~M. Generalized Gradient Approximation Made
  Simple. \emph{Phys. Rev. Lett.} \textbf{1996}, \emph{77}, 3865\relax
\mciteBstWouldAddEndPuncttrue
\mciteSetBstMidEndSepPunct{\mcitedefaultmidpunct}
{\mcitedefaultendpunct}{\mcitedefaultseppunct}\relax
\EndOfBibitem
\bibitem[Garrity \latin{et~al.}(2014)Garrity, Bennett, Rabe, and
  Vanderbilt]{garrity2014pseudopotentials}
Garrity,~K.~F.; Bennett,~J.~W.; Rabe,~K.~M.; Vanderbilt,~D. Pseudopotentials
  for High-throughput DFT Calculations. \emph{Comput. Mater. Sci.}
  \textbf{2014}, \emph{81}, 446--452\relax
\mciteBstWouldAddEndPuncttrue
\mciteSetBstMidEndSepPunct{\mcitedefaultmidpunct}
{\mcitedefaultendpunct}{\mcitedefaultseppunct}\relax
\EndOfBibitem
\bibitem[Brown \latin{et~al.}(2016)Brown, Sundararaman, Narang, Goddard~III,
  and Atwater]{brown2016ab}
Brown,~A.~M.; Sundararaman,~R.; Narang,~P.; Goddard~III,~W.~A.; Atwater,~H.~A.
  Ab Initio Phonon Coupling and Optical Response of Hot Electrons in Plasmonic
  Metals. \emph{Phys. Rev. B} \textbf{2016}, \emph{94}, 075120\relax
\mciteBstWouldAddEndPuncttrue
\mciteSetBstMidEndSepPunct{\mcitedefaultmidpunct}
{\mcitedefaultendpunct}{\mcitedefaultseppunct}\relax
\EndOfBibitem
\bibitem[Habib \latin{et~al.}(2018)Habib, Florio, and
  Sundararaman]{habib2018hot}
Habib,~A.; Florio,~F.; Sundararaman,~R. Hot Carrier Dynamics in Plasmonic
  Transition Metal Nitrides. \emph{J. Opt.} \textbf{2018}, \emph{20},
  064001\relax
\mciteBstWouldAddEndPuncttrue
\mciteSetBstMidEndSepPunct{\mcitedefaultmidpunct}
{\mcitedefaultendpunct}{\mcitedefaultseppunct}\relax
\EndOfBibitem
\end{mcitethebibliography}
 \end{document}